\documentclass[11pt,a4paper]{article}
\usepackage{amsmath,amssymb,graphicx,cite}

\def\inn{\!\cdot\!}

\makeatletter\def\@maketitle{%
  \newpage
  \null
  \vskip 2em%
  \begin{flushleft}%
  \let \footnote \thanks
    {\LARGE \@title \par}%
    \vskip 2.7em%
    {\normalsize
      \lineskip .5em%
      \hspace*{3.5em}\begin{tabular}[t]{l}%
        \@author
      \end{tabular}\par}%
  \end{flushleft}%
  \par
  \vskip 5em}\makeatother

\renewenvironment{abstract}{%
        \noindent\normalsize
          {\bfseries \large\MakeUppercase{\abstractname.}\quad}}
      {\vspace{1em} \\}

\begin{document}

\setcounter{figure}{0}
\setcounter{table}{0}
\setcounter{footnote}{0}
\setcounter{equation}{0}

\title{SOME PROPERTIES OF EMISSION COORDINATES}
\author{J.M. POZO\\[3pt]
	SYRTE, Observatoire de Paris -- CNRS\\
	61, Avenue de l'Observatoire. F-75014 Paris, France\\[3pt]
	Departament de F\'{\i}sica Fonamental, Universitat de Barcelona\\ 
	Mart\'{\i} i Franqu\`es, 1. E-08028 Barcelona, Spain\\[2pt]
	e-mail: jose-maria.pozo@obspm.fr
	}

\maketitle

\begin{abstract}
4 emitters broadcasting an increasing electromagnetic signal generate a system of relativistic coordinates for the space-time, called emission coordinates. Their physical realization requires an apparatus similar to the one of the Global Navigation Satellite Systems (GNSS).
Several relativistic corrections are utilized for
the current precisions, but the GNSS are conceived as classical (Newtonian) systems, which has deep implications in the way of operating them. The study of emission coordinates is an essential step in order to develop a fully relativistic theory of positioning systems. This talk presents some properties of emission coordinates. In particular, we characterize how any observer sees a configuration of satellites giving a degenerated system and show
that the trajectories of the satellites select a unique privileged observer at each point and, for any observer, a set of 3 orthogonal spatial axes.
\end{abstract}

\section{INTRODUCTION}

Practically all experiments in general relativity and all the uses of relativity in any application are done from a classical (Newtonian) conceptual framework. In this framework, the ``relativistic effects" are added with the same status as any non-desired perturbation (gravitational influence of other planets, effects of the atmosphere ...). This is made with corrections of first order, coming from general relativity when compared with classical mechanics, in weak gravitational fields and with small velocities (PostNewtonian formalism). Typically, this is what is done in the Global Navigation Satellite Systems (GNSS), as the GPS or the future GALILEO, where the corrections from both special and general relativity cannot be neglected.

This approach is perfectly justified from a practical and numerical point of view. However, if the resolutions are increased (more accurate clocks), it will be necessary to consider corrections of third or higher order. Then, it can be wondered if it would not be more convenient to change the framework to an exact formulation in general relativity. This would imply to abandon the classical framework. Obviously, this is a jump with many implications and difficulties of many different kinds: from technical to sociological. The first one is that such a relativistic theory of positioning systems has not been developed. Our project is aimed to develop a theory of positioning systems in the framework of general relativity. This is a long term project which is still in a state of theoretical construction.

Four emitters broadcasting an increasing electromagnetic signal generate a system of space-time coordinates, the so called {\em emission coordinates}. The most natural case is the one in which the emitted signal is the proper time of the emitter. The emission coordinates of an event (the 4 signals received) can be immediately known by this event, thus they constitute an immediate relativistic positioning system. Its physical construction is very similar to the one realized by the GNSS, where the emitters are satellites. But the way of operating and conceiving the positioning systems is very different. This is reflected in the fact that the 4 signals emitted by the satellites
are not considered as primary space-time coordinates, but the satellites are used as mere beacons to obtain separately the time and the position in some predefined coordinates for the Earth. The study of emission coordinates is aimed to develop a fully relativistic theory of positioning and reference systems. A complete theory should be able to substitute the nowadays classical perturbative approach to the satellite navigation and to provide a different framework for experimental tests of general relativity.

Emission coordinates and the associated positioning systems has been extensively studied in 2-dimensional space-times, where several strong analytic results have been obtained \cite{Coll2001,Coll2002}. Unfortunately, they are not trivially generalizable for the real case of 4-dimensional space-times, where a more deep study is needed. But, some very interesting global and local properties have been already obtained for 3 and 4 dimensions \cite{Derrick1981,CollMorales1991, Rovelli2002,Blagojevic2002,PozoColl2005}. In this work we briefly explain some local properties of 4-dimensional emission coordinates.

\section{EMISSION COORDINATES. A DEFINING DESCRIPTION}

The first ingredient in our approach to positioning systems is the use of the 4 electromagnetic signals emitted by each set of 4 satellites, directly as coordinates for the domain of interest of the space-time. This type of coordinates are called {\em emission coordinates}. The usual treatments of any relativistic problem, takes a somehow classical description of the space-time, using one time-like coordinate which defines the instantaneous 3-dimensional spaces (synchronization) and 3 space-like coordinates to coordinate this sequence of spaces. This description is very adapted to our intuition about space and time. However, when relativity is not negligible, the needed synchronization is a convention which is not absolutely defined. For inertial observers in flat space-time we can use the Einstein convention, which is dependent on the observer chosen. For general observers, even in flat space-time, this is still worse since no standard synchronization is well defined.

\begin{figure}[htb]
\centerline{
\includegraphics[width=0.45\textwidth]{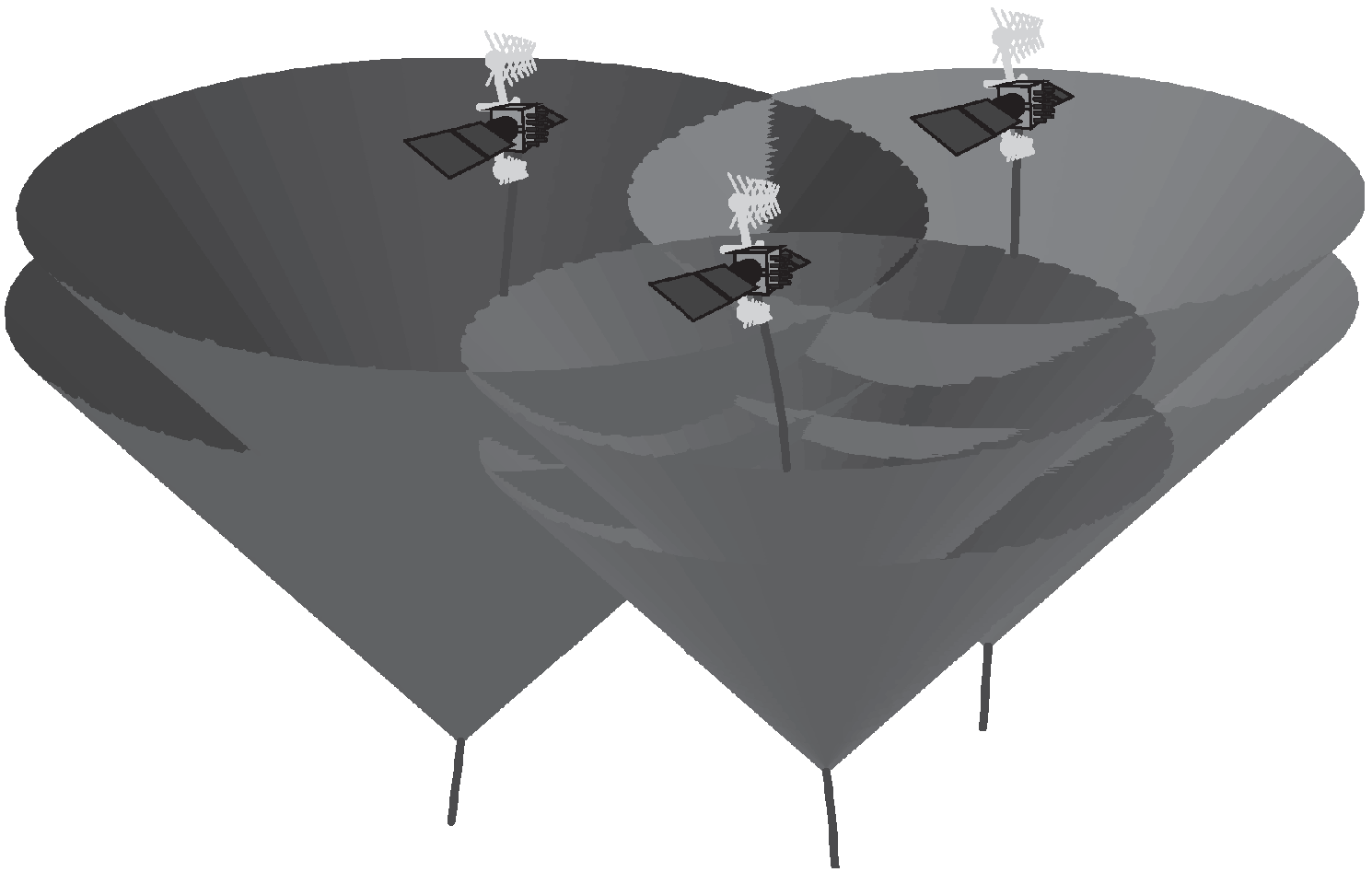}\hspace*{0.07\textwidth}
\includegraphics[width=0.26\textwidth]{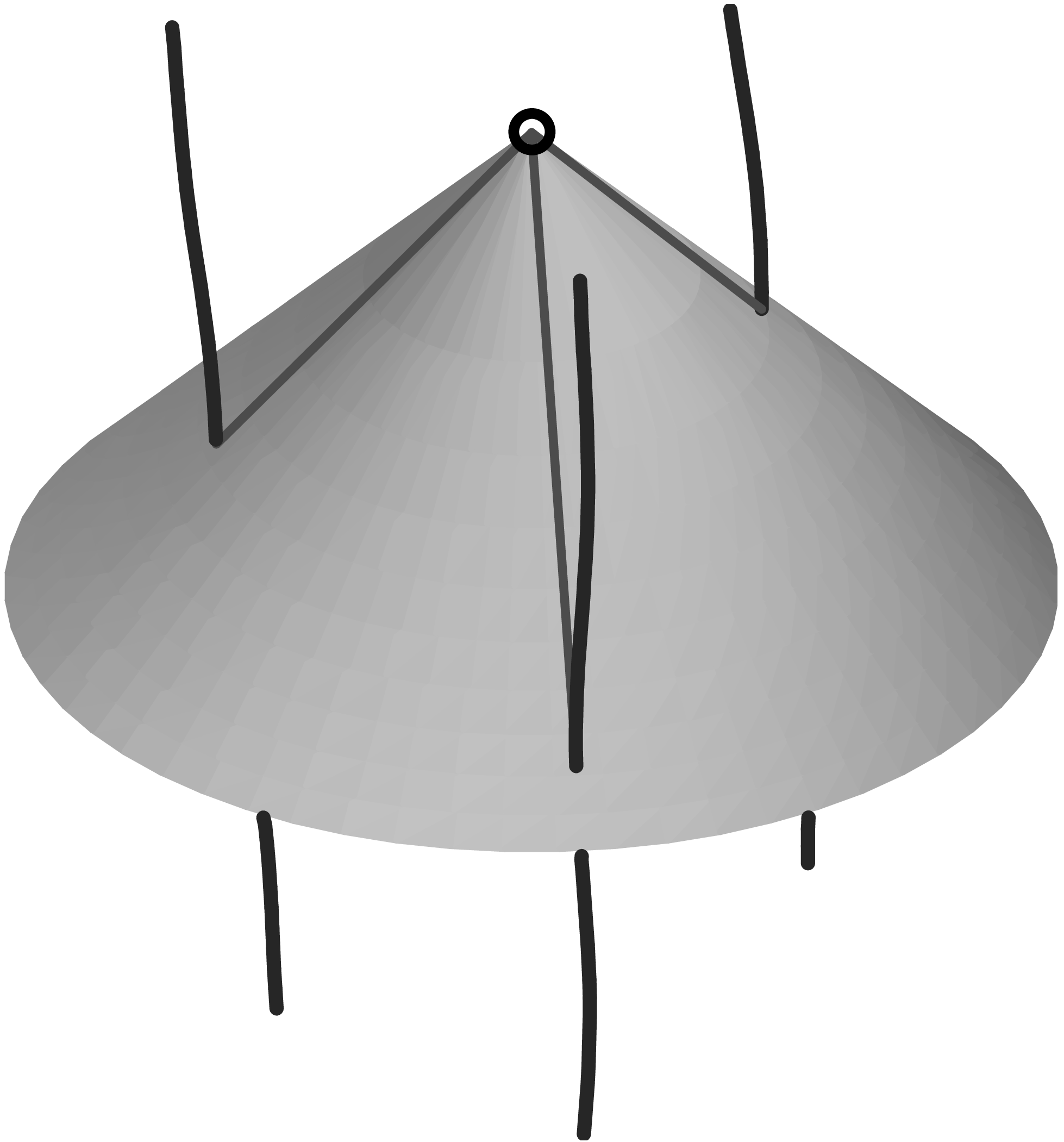}}
\caption{\small Representation of 3 emitters in a 3-dimensional space-time (time vertical). The lines are the space-time trajectories (world-lines) of the emitters. The left figure shows the space-time surfaces visited by each value of the signals emitted (future light-cones), which defines the grid of emission coordinates. The right figure shows the past light-cone of an event. Its intersection with the trajectories of the emitters gives the emission coordinates of this event.}
\end{figure}

The 4 families of hypersurfaces defining emission coordinates are light-like. This means that the natural covectors $\{{\rm d}\tau^1,{\rm d}\tau^2,{\rm d}\tau^3,{\rm d}\tau^4\}$ are light-like:
$
	{\rm d}\tau^A\inn{\rm d}\tau^A=0
$.
In fact the metrically associated vectors
$\vec\ell^A=g^*({\rm d}\tau^A)$ (or with index notation, $(\vec\ell^A)^\mu=g^{\mu\nu}({\rm d}\tau^A)_\nu$)
are the directions defined by the light-like geodesics followed by the signals (rays). Observe that this class of coordinates is radically different from the usual decomposition into space and time, where we have 1 time-like coordinate and 3 space-like coordinates.

\section{PROPERTIES OF THE METRIC}

The light-like nature of the covectors implies that the contravariant metric in emission coordinates has vanishing diagonal elements:
\[
	(g^{AB})=\begin{pmatrix}
		0& g^{12}& g^{13}& g^{14} \\
		g^{12}& 0& g^{23}& g^{24} \\
		g^{13}& g^{23}& 0& g^{34} \\
		g^{14}& g^{24}& g^{34}& 0
	\end{pmatrix}
\]
And, since the covectors ${\rm d}\tau^A$ are future oriented, the extra-diagonal elements are all positive $g^{AB}>0$. Besides, the Lorentzian signature of the space time implies the triangular inequalities
\[
	A<B+C, \quad B<A+C,\quad C<A+B
	\quad\text{where}\quad
	A\equiv\sqrt{g^{14}g^{23}},\quad B\equiv\sqrt{g^{24}g^{13}},\quad 
	C\equiv\sqrt{g^{34}g^{12}} \,.
\]
The determinant of the metric can be factorized into
\[
	|g^{AB}| =(A+B+C)(A-B-C)(B-A-C)(C-A-B) \,.
\]

Given an observer $u$ (that is, a field of unit time-like vectors representing the {\em 4-velocity} of a local laboratory) the light-like vector $\vec\ell^A$ is split into
\[
		\label{eq:RayInSpace&Time}
	\vec\ell^A=\nu^A(u+\hat n^A),
\]
where $\nu^A$ is a positive scalar representing the {\it frequency} of 
the signal seen by the observer, and $\hat n^A$ is a space-like unitary vector, 
$\hat n^A\cdot\hat n^A=-1$, representing the direction toward which 
the observer sees the propagation of the signal in its space.

The angle $\theta^{AB}$ between the directions
$\hat n^A$ and $\hat n^B$ of two different signals (the apparent angle between the two signals) are given by
\[
	c^{AB}\equiv\cos\theta^{AB}=-\hat n^A\cdot\hat n^B \ .
\]

\noindent {\bfseries Result}\ \ {\itshape Given an arbitrary observer at an event, 4 light-like directions, $d\tau^A$, are linearly dependent at this event if and only if the observer sees the apparent sources of the signals $\tau^A$ arranged in a circle in its celestial sphere.}

\section{THE INTRINSIC SPLITTING OF THE METRIC AND THE CENTRAL OBSERVER}

Given a system of emission coordinates we can ask if there exists some observers who see the constellation of satellites arranged in some special configuration.
For instance, in 3 dimensions we can ask for an observer who see the 3 satellites in his celestial {\em circumference} with all the angles equal: $\theta^{12}=\theta^{13}=\theta^{23}=2\pi/3$. This property can be generalized to 4 dimensions by asking for the 6 angles $\theta^{AB}$ between the 4 emitters to be equal. This would correspond to an observer who would see the 4 emitters arranged in a regular tetrahedron in its celestial sphere. This property is too restrictive. A different generalization is to ask for the 4 solid angles defined by the trihedral of each 3 emitters to be equal: $\theta^{123}=\theta^{124}=\theta^{134}=\theta^{234}=\pi$. This corresponds to an observer who sees the 4 emitters in an {\em equifacial tetrahedron} in its celestial sphere. This property can be equivalently characterized by asking for the angle between each pair of directions to be equal to the angle formed by the complementary pair: $\theta^{12}=\theta^{34}$, $\theta^{13}=\theta^{24}$ and $\theta^{23}=\theta^{14}$. An observer seeing the four emitters in this configuration will be called a {\em Central observer}. Let us remark that (although called central) this property does not selects a position or an event in the space-time, but a velocity at each event.

\noindent {\bfseries Result}\ \ {\itshape Given any system of emission coordinates the Central observer exists and is unique.}

An equifacial tetrahedron defines 3 orthogonal axes in the space. This implies that the 4 trajectories of the emitters also select 3 orthogonal spatial directions, orthogonal to the central observer. In addition, since any two observers define a pure boost that relate both, by applying this boost to the spatial axes we obtain the following:

\noindent {\bfseries Result}\ \ {\itshape For any observer, the trajectories of the 4 emitters select 3 privileged orthogonal axis in its space.}

\begin{figure}[htb]
\centerline{
\includegraphics[width=0.21\textwidth]{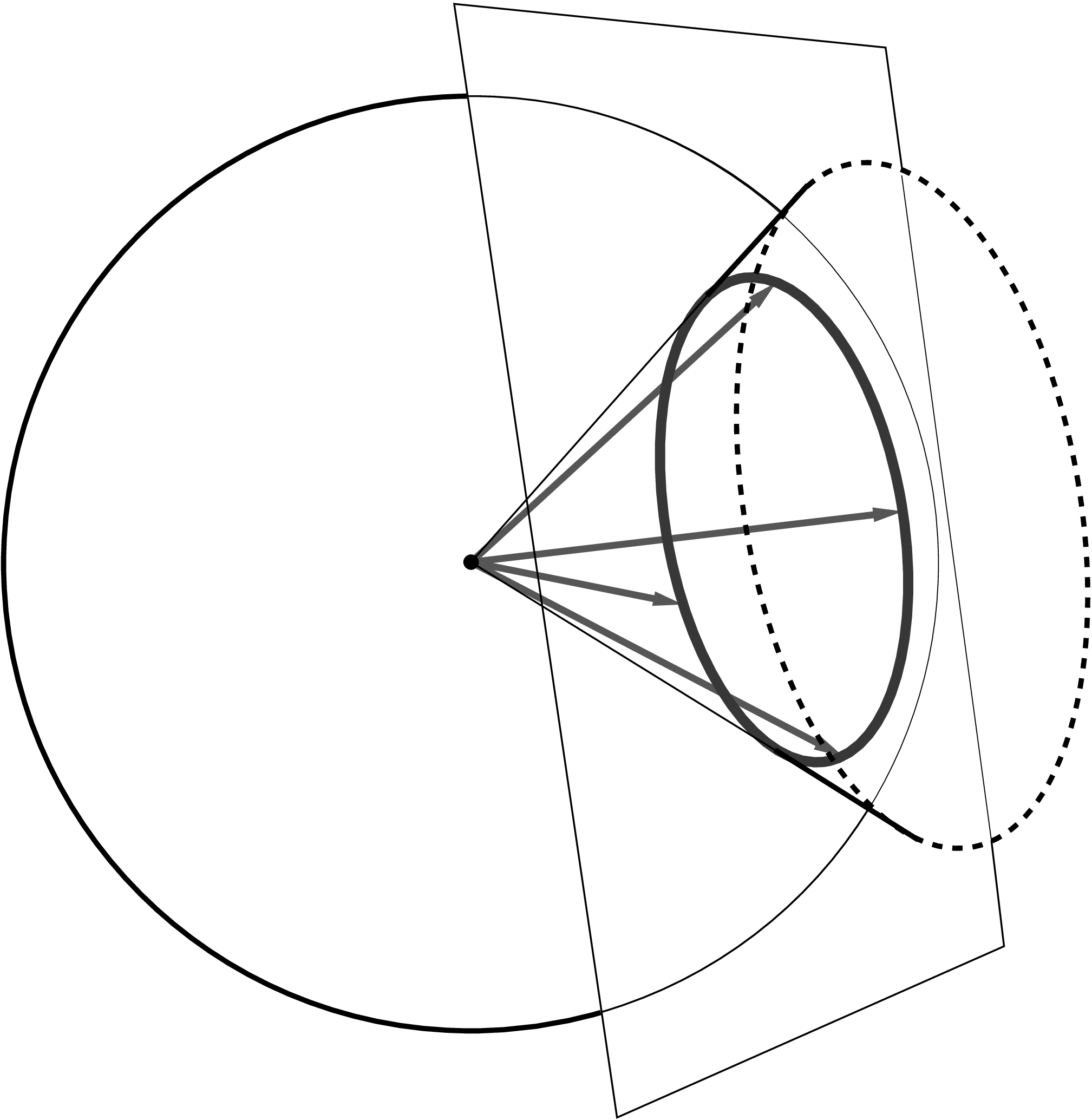}\hspace*{0.13\textwidth}
\includegraphics[width=0.26\textwidth]{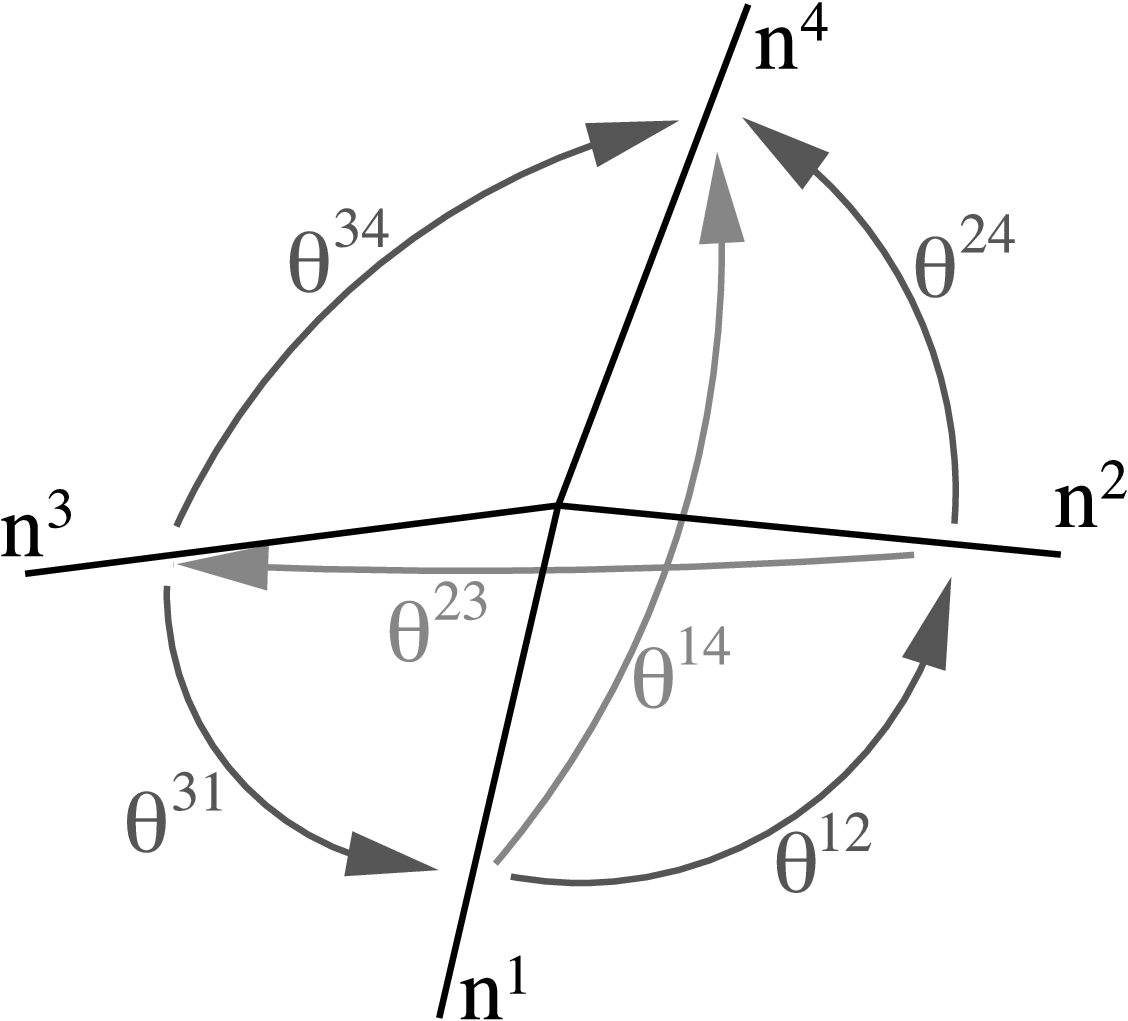}}
\caption{\small The left figure represents 4 points in the celestial sphere of an observer which lie in a unique circle (the directions lie in a cone). The system of emission coordinates is degenerate at the events where any observer see the 4 emitters in this configuration. The right figure represents 4 directions and the six planar angles between them. 
}
\end{figure}

This result is intimately related with the existence of the following splitting of the metric:
\[
	(g^{AB})=\begin{pmatrix}
		\mu^1 & 0 & 0 & 0 \\
		0 & \mu^2 & 0 & 0 \\
		0 & 0 & \mu^3 & 0 \\
		0 & 0 & 0 & \mu^4
	\end{pmatrix}
	\begin{pmatrix}
		0 & \hat C & \hat B & \hat A \\
		\hat C & 0 & \hat A & \hat B \\
		\hat B & \hat A & 0 & \hat C \\
		\hat A & \hat B & \hat C & 0
	\end{pmatrix}
	\begin{pmatrix}
		\mu^1 & 0 & 0 & 0 \\
		0 & \mu^2 & 0 & 0 \\
		0 & 0 & \mu^3 & 0 \\
		0 & 0 & 0 & \mu^4
	\end{pmatrix}
\]
Here, the 6 degrees of freedom of the metric are split into two types of parameters, which behave in clearly different ways when the series broadcasted by the emitters is changed (that is, when the clocks on board the satellites are modified):

\begin{itemize}

\item 
4 parameters $\{\mu^A\}$ which scale, each of them, linearly with the series broadcasted by the corresponding satellite and are independent of the others.

\item 
3 parameters (2 degrees of freedom) $\hat A,\hat B,\hat C$ which are invariant respect to the change of the series broadcasted. They only dependent on the satellite world-lines.

\end{itemize}

This splitting is a promising tool in order to study the consequences of the derive of the clocks, the election of different corrections for this, and the possible role that some arbitrary or geometrically justified synchronization could have in the positioning system.

\end{document}